Astronomy
&
Astrophysics

# Dynamical friction of a massive black hole in a turbulent gaseous medium

Sandrine Lescaudron[1], Yohan Dubois[1], Ricarda S. Beckmann[1,2], and Marta Volonteri[1]

[1] Sorbonne Université, CNRS, UMR 7095, Institut d'Astrophysique de Paris, 98 Bis bd Arago, 75014 Paris, France
e-mail: sandrine.lescaudron@iap.fr
[2] Institute of Astronomy and Kavli Institute for Cosmology, University of Cambridge, Madingley Road, Cambridge CB3 0HA, UK

**ABSTRACT**

The orbital decay of massive black holes in galaxies in the aftermath of mergers is at the heart of whether massive black holes successfully pair and merge, leading to emission of low-frequency gravitational waves. The role of dynamical friction sourced from the gas distribution has been uncertain because many analytical and numerical studies have either focussed on a homogeneous medium or have not reached resolutions below the scales relevant to the problem, namely the Bondi-Hoyle-Lyttleton radius. We performed numerical simulations of a massive black hole moving in a turbulent medium in order to study dynamical friction from turbulent gas. We find that the black hole slows down to the sound speed, rather than the turbulent speed, and that the orbital decay is well captured if the Bondi-Hoyle-Lyttleton radius is resolved with at least five resolution elements. We find that the larger the turbulent eddies, the larger the scatter in dynamical friction magnitude, because of the stochastic nature of the problem, and also because of the larger over- and under-densities encountered by the black hole along its trajectory. Compared to the classic solution in a homogeneous medium, the magnitude of the force depends more weakly on the Mach number, and dynamical friction is overall more efficient for high Mach numbers, but less efficient towards and at the transonic regime.

## 1. Introduction

The evolution of black holes (BHs), their ability to accrete surrounding matter, or to undergo mergers is directly linked to their dynamical evolution, their position in the galaxy, and their velocity relative to the ambient medium. In general, BHs move in the potential of their host galaxy. More locally, they exchange momentum and energy with the interstellar medium and the stars through dynamical friction.

Dynamical friction can be defined as the interaction of a massive object with its own gravitational wake. A massive object moving through lighter bodies or a gaseous medium attracts the surrounding matter, creating an over-density (wake) behind it. The wake in turn exerts a gravitational force on the massive object and slows it down. In this way, momentum and kinetic energy are transferred from the massive object to its over-dense wake.

Dynamical friction is an important force in many different contexts, including the evolution of BH binaries (Yu 2002; Dotti et al. 2007; Mayer 2013; Dosopoulou & Antonini 2017; Li et al. 2020b) and X-ray binaries (Iben & Livio 1993; Hurley et al. 2002), the orbital evolution of black holes in galaxies (Volonteri & Perna 2005; Bellovary et al. 2010; Tremmel et al. 2017), the orbital decay of a galaxy satellite in a galaxy cluster (Colpi et al. 1999; Fujii et al. 2006; Ogiya & Burkert 2016), among others. The detection of gravitational waves from supermassive BHs, which is possible with the upcoming Laser Interferometer Space Antenna (LISA; Amaro-Seoane et al. 2017) and through pulsar timing arrays (Jenet et al. 2004, 2005), has made the orbital decay of supermassive BHs in a galaxy merger remnant a topic of special interest, as one needs to determine under which conditions supermassive BHs could coalesce and emit gravitational waves in less than a Hubble time. Furthermore, supermassive BHs can accrete efficiently and for a prolonged time from their surroundings, and exert feedback on the host galaxy, which is considered of paramount importance to regulate the baryon content in halos and galaxies, and to set up diverse galaxy properties (e.g. Granato et al. 2004; Di Matteo et al. 2005; Croton et al. 2006; Bower et al. 2006; Sijacki et al. 2007; Dubois et al. 2010, 2012, 2016; Beckmann et al. 2017).

There is increasing theoretical support for wandering massive BHs within low-mass galaxies (Bellovary et al. 2019; Pfister et al. 2019; Lapiner et al. 2021; Ricarte et al. 2021), which might correspond to spatially offset active galactic nuclei as observed with radio emission in a number of dwarfs (Reines et al. 2020). This population of wandering BHs originate from galaxies tidally perturbed by a merger (Bellovary et al. 2021), can be scattered within galaxies by their multiphase structure (Pfister et al. 2019), or can be produced by gravitational recoil (Blecha et al. 2011). Therefore, accurately modelling the BH dynamics, which should consider the effect of dynamical friction, is key to BH mass growth and feedback (Lapiner et al. 2021; Bahé et al. 2022), their sinking into galaxies (Tremmel et al. 2017; Bartlett et al. 2021; Ma et al. 2021), and, hence, of their merger rate (Volonteri et al. 2020; Li et al. 2020a; Barausse et al. 2020; Chen et al. 2022; Kunyang et al. 2022).

Dynamical friction was first described analytically by Chandrasekhar (1943) who developed the theory for a collisionless medium, such as stars or dark matter particles. Then Ostriker (1999) proposed an analytical solution for a uniform

Lescaudron, S., et al.: A&A 674, A217 (2023)gaseous medium using time-dependent linear perturbation theory. Ostriker (1999) also showed, by comparing her work to the work of Chandrasekhar, that the gaseous dynamical friction is less efficient than the collisionless dynamical friction in the subsonic regime but much more efficient in the transonic regime where it sharply peaks. For large Mach number dynamical friction force is the same in a gaseous or collisionless medium. Recent insights have shown that dynamical friction from gas and stars both play a crucial role in allowing orbits of BH to decay in galaxies (Pfister et al. 2019; Chen et al. 2022), with gas dynamical friction particularly important at high redshift.The non-linear evolution of dynamical friction has been studied using idealised numerical experiments. Early work was done by Ruffert & Arnett (1994) and Ruffert (1996) who studied the evolution of the wake and the resulting force on the BH using uniform background gas density and sound speed. Like most work in the field, these experiments were conducted in the frame of the BH, with the background gas moving at a fixed Mach number at all times. During this work, it became apparent that, as predicted in analytic models (Cowie 1977), wakes are unstable, a phenomenon further explored in Foglizzo et al. (2005). Since then, studies have investigated the impact of various physical phenomena on dynamical friction, such as for example density gradients (MacLeod & Ramirez-Ruiz 2015) or the impact of BH radiative feedback on wake structure (Park & Bogdanović 2017, 2019).

As most of the force onto the BH is produced by density perturbations very close to the BH, resolving dynamical friction in large-scale cosmological simulations is numerically very expensive. Instead, it can be included in simulations using analytic sub-grid models (Dubois et al. 2013, 2021; Tremmel et al. 2015; Pfister et al. 2019; Ni et al. 2022). While not without its challenges (Korol et al. 2016; Beckmann et al. 2018; Morton et al. 2021), this method has allowed for crucial improvements in reliably calculating the orbits of massive BHs in galaxies.

In this paper we study dynamical friction from a gaseous medium. The work presented here bridges the gap between isolated numerical experiments and full galaxy scale simulations to investigate the impact of small-scale turbulence on the dynamical friction force. The interstellar medium being turbulent, the conditions adopted here are more realistic than the thoroughly investigated, but idealised, case of a homogeneous gaseous medium.

The paper is organised as follows. In Sect. 2 we recall the theoretical background and provide definitions and modifications for our study. In Sect. 3 we introduce the numerical models. In Sect. 4 we describe our simulations and their initial conditions. Results are presented in Sect. 5. Section 6 is devoted to discussion and conclusions.

## 2. Theory of dynamical friction

The theory of dynamical friction in a gaseous medium was first based on linear theory under the assumption of a steady state. In the supersonic case, Dokuchaev (1964), Ruderman & Spiegel (1971), and Rephaeli & Salpeter (1980) obtained the following formula for the dynamical friction force exerted on a massive perturber:

$$F_{SS} = \frac{4\pi(GM_{BH})^2 \rho_0}{v_{rel}^2} \ln\left(\frac{r_{max}}{r_{min}}\right) \tag{1}$$

where $G$ is the gravitational constant, $M_{BH}$ is the mass of the perturber, $\rho_0$ the background gas density, $v_{rel}$ the relative velocity of the perturber with respect to the background gas, and $r_{min}$ and $r_{max}$ are respectively the size of the perturber and the characteristic size of the surrounding medium. In many situations, $r_{min}$ and $r_{max}$ are poorly defined but due to the logarithm, large variations in the two characteristic sizes leads to a small difference in the overall force.

In the subsonic case, Rephaeli & Salpeter (1980) argued that the dynamical friction force is null due to the symmetrical distribution of the gas around the massive object. However, this hypothesis shows a discontinuity between the supersonic and subsonic regimes, when the dynamical friction force, maximal just above Mach number suddenly goes to zero. This non-physical result led Ostriker (1999) to review the problem under a time-dependent assumption.

The time-dependent theory allows to calculate a non-zero dynamical friction force in the subsonic regime:

$$F_{Sub} = \frac{4\pi(GM_{BH})^2 \rho_0}{v_{rel}^2} \left[\frac{1}{2}\ln\left(\frac{1+\mathcal{M}}{1-\mathcal{M}}\right) - \mathcal{M}\right], \tag{2}$$

and to take into account the development of the wake in the supersonic regime:

$$F_{Sup} = \frac{4\pi(GM_{BH})^2 \rho_0}{v_{rel}^2} \left[\frac{1}{2}\ln\left(1 - \frac{1}{\mathcal{M}^2}\right) + \ln\frac{r_{max}}{r_{min}}\right], \tag{3}$$

where $\mathcal{M} = v_{rel}/c_s$ is the Mach number, and $c_s$ is the sound speed. The analytic force is therefore discontinuous at the sonic point, but a linear interpolation can be used to link the two regimes.

According to Ostriker (1999), the characteristic size of the medium $r_{max}$ is equal to the size of the wake, $r_{wake}$. This, in turn can be calculated from $v_{rel}t$ (where $t$ is the time after which the perturber has started to exert its force on the background medium) under the assumptions that $(v_{rel} - c_s)t$ is bigger than the perturber size $r_{min}$ and that $(v_{rel} + c_s)t$ is smaller than the surrounding medium size $r_{max}$.

For the turbulent medium we are interested in, some additional definitions are needed. In this paper, we use two different definitions of the Mach number. Firstly the classic Mach number $\mathcal{M} = v_{rel}/c_s$. Throughout the text, categorisations such as 'supersonic' ($\mathcal{M} > 1$), 'transonic' ($\mathcal{M} \sim 1$), and 'subsonic' ($\mathcal{M} < 1$) refer to the classic Mach number. Secondly a turbulent Mach number $\mathcal{M}_{turb} = v_{rel}/v_{eff}$ where $v_{eff} = \sqrt{c_s^2 + v_{rms}^2}$ is an effective gas velocity that captures the contribution of the gas sound speed and of the turbulent velocity ($v_{rms}$ is the root mean square velocity of the gas).

For a turbulent medium, we expect the time it takes for the stirring of the turbulence to erase the wake over-density to determine the typical length of the wake, $r_{wake}$. The turbulence will erase the wake on a characteristic timescale $t_{cross} = r_{BHL}/v_{eff}$, which is equal to the time taken by the effective turbulent velocity $v_{eff}$ to cross the width of the wake, taken equal to the Bondi-Hoyle-Lyttleton (BHL) radius $r_{BHL}$, defined as:

$$r_{BHL} = \frac{2GM_{BH}}{v_{rel}^2 + v_{eff}^2} \tag{4}$$

$$= 0.43 \frac{M_{BH}}{10^4\,M_\odot} \left[\left(\frac{v_{rel}}{10\,\mathrm{km\,s}^{-1}}\right)^2 + \left(\frac{v_{eff}}{10\,\mathrm{km\,s}^{-1}}\right)^2\right]^{-1}\,\mathrm{pc}.$$

Hence, the length of the wake is expected to simply be $r_{wake} = t_{cross}v_{rel} = r_{BHL}\mathcal{M}_{turb}$ when fully established. Taking early times into account, we have:

$$r_{wake} = \begin{cases} v_{rel}t & \text{when } t < t_{cross} \\ r_{BHL}\mathcal{M}_{turb} & \text{otherwise.} \end{cases} \tag{5}$$

A217, page 2 of 14

gaseous medium using time-dependent linear perturbation theory. Ostriker (1999) also showed, by comparing her work to the work of Chandrasekhar, that the gaseous dynamical friction is less efficient than the collisionless dynamical friction in the subsonic regime but much more efficient in the transonic regime where it sharply peaks. For large Mach number dynamical friction force is the same in a gaseous or collisionless medium. Recent insights have shown that dynamical friction from gas and stars both play a crucial role in allowing orbits of BH to decay in galaxies (Pfister et al. 2019; Chen et al. 2022), with gas dynamical friction particularly important at high redshift.

The non-linear evolution of dynamical friction has been studied using idealised numerical experiments. Early work was done by Ruffert & Arnett (1994) and Ruffert (1996) who studied the evolution of the wake and the resulting force on the BH using uniform background gas density and sound speed. Like most work in the field, these experiments were conducted in the frame of the BH, with the background gas moving at a fixed Mach number at all times. During this work, it became apparent that, as predicted in analytic models (Cowie 1977), wakes are unstable, a phenomenon further explored in Foglizzo et al. (2005). Since then, studies have investigated the impact of various physical phenomena on dynamical friction, such as for example density gradients (MacLeod & Ramirez-Ruiz 2015) or the impact of BH radiative feedback on wake structure (Park & Bogdanović 2017, 2019).

As most of the force onto the BH is produced by density perturbations very close to the BH, resolving dynamical friction in large-scale cosmological simulations is numerically very expensive. Instead, it can be included in simulations using analytic sub-grid models (Dubois et al. 2013, 2021; Tremmel et al. 2015; Pfister et al. 2019; Ni et al. 2022). While not without its challenges (Korol et al. 2016; Beckmann et al. 2018; Morton et al. 2021), this method has allowed for crucial improvements in reliably calculating the orbits of massive BHs in galaxies.

In this paper we study dynamical friction from a gaseous medium. The work presented here bridges the gap between isolated numerical experiments and full galaxy scale simulations to investigate the impact of small-scale turbulence on the dynamical friction force. The interstellar medium being turbulent, the conditions adopted here are more realistic than the thoroughly investigated, but idealised, case of a homogeneous gaseous medium.

The paper is organised as follows. In Sect. 2 we recall the theoretical background and provide definitions and modifications for our study. In Sect. 3 we introduce the numerical models. In Sect. 4 we describe our simulations and their initial conditions. Results are presented in Sect. 5. Section 6 is devoted to discussion and conclusions.

## 2. Theory of dynamical friction

The theory of dynamical friction in a gaseous medium was first based on linear theory under the assumption of a steady state. In the supersonic case, Dokuchaev (1964), Ruderman & Spiegel (1971), and Rephaeli & Salpeter (1980) obtained the following formula for the dynamical friction force exerted on a massive perturber:

$$F_{SS} = \frac{4\pi(GM_{BH})^2 \rho_0}{v_{rel}^2} \ln\left(\frac{r_{max}}{r_{min}}\right) \tag{1}$$

where $G$ is the gravitational constant, $M_{BH}$ is the mass of the perturber, $\rho_0$ the background gas density, $v_{rel}$ the relative velocity of the perturber with respect to the background gas, and $r_{min}$ and $r_{max}$ are respectively the size of the perturber and the characteristic size of the surrounding medium. In many situations, $r_{min}$ and $r_{max}$ are poorly defined but due to the logarithm, large variations in the two characteristic sizes leads to a small difference in the overall force.

In the subsonic case, Rephaeli & Salpeter (1980) argued that the dynamical friction force is null due to the symmetrical distribution of the gas around the massive object. However, this hypothesis shows a discontinuity between the supersonic and subsonic regimes, when the dynamical friction force, maximal just above Mach number suddenly goes to zero. This non-physical result led Ostriker (1999) to review the problem under a time-dependent assumption.

The time-dependent theory allows to calculate a non-zero dynamical friction force in the subsonic regime:

$$F_{Sub} = \frac{4\pi(GM_{BH})^2 \rho_0}{v_{rel}^2} \left[\frac{1}{2}\ln\left(\frac{1+\mathcal{M}}{1-\mathcal{M}}\right) - \mathcal{M}\right], \tag{2}$$

and to take into account the development of the wake in the supersonic regime:

$$F_{Sup} = \frac{4\pi(GM_{BH})^2 \rho_0}{v_{rel}^2} \left[\frac{1}{2}\ln\left(1 - \frac{1}{\mathcal{M}^2}\right) + \ln\frac{r_{max}}{r_{min}}\right], \tag{3}$$

where $\mathcal{M} = v_{rel}/c_s$ is the Mach number, and $c_s$ is the sound speed. The analytic force is therefore discontinuous at the sonic point, but a linear interpolation can be used to link the two regimes.

According to Ostriker (1999), the characteristic size of the medium $r_{max}$ is equal to the size of the wake, $r_{wake}$. This, in turn can be calculated from $v_{rel}t$ (where $t$ is the time after which the perturber has started to exert its force on the background medium) under the assumptions that $(v_{rel} - c_s)t$ is bigger than the perturber size $r_{min}$ and that $(v_{rel} + c_s)t$ is smaller than the surrounding medium size $r_{max}$.

For the turbulent medium we are interested in, some additional definitions are needed. In this paper, we use two different definitions of the Mach number. Firstly the classic Mach number $\mathcal{M} = v_{rel}/c_s$. Throughout the text, categorisations such as 'supersonic' ($\mathcal{M} > 1$), 'transonic' ($\mathcal{M} \sim 1$), and 'subsonic' ($\mathcal{M} < 1$) refer to the classic Mach number. Secondly a turbulent Mach number $\mathcal{M}_{turb} = v_{rel}/v_{eff}$ where $v_{eff} = \sqrt{c_s^2 + v_{rms}^2}$ is an effective gas velocity that captures the contribution of the gas sound speed and of the turbulent velocity ($v_{rms}$ is the root mean square velocity of the gas).

For a turbulent medium, we expect the time it takes for the stirring of the turbulence to erase the wake over-density to determine the typical length of the wake, $r_{wake}$. The turbulence will erase the wake on a characteristic timescale $t_{cross} = r_{BHL}/v_{eff}$, which is equal to the time taken by the effective turbulent velocity $v_{eff}$ to cross the width of the wake, taken equal to the Bondi-Hoyle-Lyttleton (BHL) radius $r_{BHL}$, defined as:

$$r_{BHL} = \frac{2GM_{BH}}{v_{rel}^2 + v_{eff}^2} \tag{4}$$

$$= 0.43 \frac{M_{BH}}{10^4\,M_\odot} \left[\left(\frac{v_{rel}}{10\,\mathrm{km\,s}^{-1}}\right)^2 + \left(\frac{v_{eff}}{10\,\mathrm{km\,s}^{-1}}\right)^2\right]^{-1}\,\mathrm{pc}.$$

Hence, the length of the wake is expected to simply be $r_{wake} = t_{cross}v_{rel} = r_{BHL}\mathcal{M}_{turb}$ when fully established. Taking early times into account, we have:

$$r_{wake} = \begin{cases} v_{rel}t & \text{when } t < t_{cross} \\ r_{BHL}\mathcal{M}_{turb} & \text{otherwise.} \end{cases} \tag{5}$$

## 3. Magneto-hydrodynamics simulations with forcing turbulence

The simulations presented here were performed with the RAMSES code (Teyssier 2002), solving for ideal magneto-hydrodynamics (MHD; Fromang et al. 2006) with self-gravity. The gas evolution was obtained with a MUSCL-Hancock scheme which used a second-order Godunov method with a minmod limiter on the linear reconstruction of the conserved quantities at cell interface. The MHD flux at each cell interface was obtained with the approximate HLLD (Harten-Lax-Van Leer-Discontinuities) Riemann solver (Miyoshi & Kusano 2005). The induction equation that evolves the magnetic $\boldsymbol{B}$ field was solved with the constrained transport algorithm that guarantees $\nabla \cdot \boldsymbol{B} = 0$ at machine accuracy (Evans & Hawley 1988; Teyssier et al. 2006).

The massive perturber was evolved with a particle-mesh method using a cloud-in-cell interpolation. The total gravitational potential, also accounting for the gas contribution, was obtained with a multi-grid solver for the Poisson equation (Guillet & Teyssier 2011). Simulations presented here were run using a uniform grid, as the turbulent nature of the gas makes numerical savings from using adaptive mesh refinement very small. In addition, a uniform grid avoids errors in the Poisson solver, that can generate a non-negligible self-force at coarse-to-fine boundaries near the BH (Bleuler & Teyssier 2014; Zhu & Gnedin 2021) due to its strong gravitational potential.

We used a specific model to drive turbulence in the interstellar medium, based on the implementation done by Commerçon et al. (2019) with further details of the implementation documented in Schmidt et al. (2009) and Federrath et al. (2010). Kinetic energy was injected using the Ornstein-Uhlenbeck process (Eswaran & Pope 1988), which creates a turbulent forcing generated in Fourier space, using a combination of compressive (1/3) and solenoidal (2/3) modes. This forcing was defined by the wavenumber $k_{\text{turb}}$, the power spectrum of shape $1 - (k - k_{\text{turb}})^2$, and an ad hoc boost factor of the driving force, which sets the amplitude of the turbulent velocity. In the following set of simulations we set the boost factor to a constant value for any choice of central $k_{\text{turb}}$, which turns into a smaller rms velocity for larger $k_{\text{turb}}$. The typical range of values of rms velocities experienced in these simulation were within $v_{\text{rmx}} = 2-4.6 \text{ km s}^{-1}$ (for $k_{\text{turb}} = 16$ to 4 respectively). The turbulence wavenumber $k_{\text{turb}}$, referred to as $k$ in the following, is a fraction of the box size and represents the mean size of turbulent eddies. The auto-correlation time was set to 0.6 Myr.

The gas could cool down to temperatures as low as 10 K using the cooling function $\mathcal{L}(\rho, T)$ from Sutherland & Dopita (1993) for temperatures above $10^4$ K and the cooling due to carbon, oxygen, and hydrogen for temperatures below $10^4$ K. We assumed solar metallicity and a typical Milky Way UV flux for ionisation counts. The gas was a perfect gas with an adiabatic index $\gamma = 5/3$ and a mean molecular weight $\mu = 1.4$. We neglected any gas accretion onto the BH, and hence its feedback, as well as star formation and feedback from stars occurring in the interstellar medium.

## 4. Simulations

### 4.1. General setup

In order to study the effect of the multi-phase gas structure seeded by the turbulence we have chosen to set an initial gas density of $\rho_0 = 10 \text{ cm}^{-3}$, within the typical range of the neutral medium so that the gas got thermally unstable and formed structures. The turbulence energy injection provided an additional heating leading to a temperature equilibrium about $T \simeq 350$ K with a standard deviation of $\sigma_T \simeq 345$ K and a sound speed about $c_s \simeq 1.2 \text{ km s}^{-1}$ with a standard deviation of $\sigma_{c_s} \simeq 0.6 \text{ km s}^{-1}$. The magnetic field at equilibrium was about $B \simeq 2.4 \mu$G, corresponding to an Alfvén velocity about $v_A \simeq 1.4 \text{ km s}^{-1}$. In our simulations, the Alfvénic Mach number ($\mathcal{M}_A = v_{\text{rms}}/v_A$) remained above one, that is in the supersonic or nearly transonic regime. Although magnetic fields can play a significant role in the anisotropic shape of the turbulence (e.g. Goldreich & Sridhar 1995; Cho & Vishniac 2000; Beattie & Federrath 2020), we deferred the investigation of its effect on dynamical friction to future work.

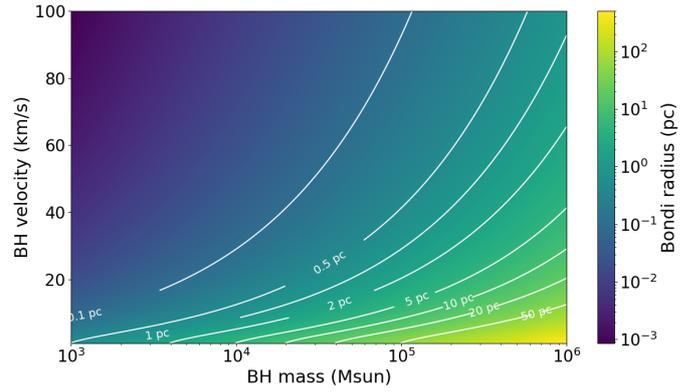

**Fig. 1.** BHL radius evolution with BH mass and velocity. The gas effective velocity considered is that of the reference case, $v_{\text{eff}} = 4 \text{ km s}^{-1}$.

In idealised simulations, the BH mass and velocity cannot be chosen arbitrarily as there are a series of constraints that must be fulfilled to combine sufficient resolution with a reasonable computational time, within a physically interesting set-up.

The resolution was set by the requirement to resolve the BHL radius $r_{\text{BHL}}$. Resolving $r_{\text{BHL}}$ was necessary to capture the gas overdensity formed by the BH gravitational potential, and, hence, the dynamical friction force exerted by the gas onto the BH (Beckmann et al. 2018, and references therein).

Furthermore, we had to take into account that $r_{\text{BHL}}$ increases over time as the BH slows down. A maximum value is reached for $r_{\text{BHL}}$ when the BH velocity is similar to the effective gas velocity. To study the problem presented here, $r_{\text{BHL}}$ must be kept at all times between a few times larger than the resolution and a few times smaller than the box size. Figure 1 shows the value of $r_{\text{BHL}}$ for our reference case for the gas properties and for different values of the BH mass and BH relative velocity. Based on this, the initial BH velocity was set to about 3 times that of the gas. This allowed us to study a range of Mach numbers while avoiding large variations of $r_{\text{BHL}}$ over time.

Given the typical $v_{\text{eff}}$ of the gas distribution, this series of constraints led to $10^4 \, M_\odot$ for the BH mass and 40 pc for the box size as acceptable values. To avoid excessive computational times we had also calculated the expected theoretical dynamical friction time, and we had checked the BH-to-gas mass ratio and the Jeans mass to avoid gas collapse, but these were less restrictive parameters. For the Jeans mass calculation we noted that the relevant velocity to consider is the effective velocity $v_{\text{eff}} \geq 3.3 \text{ km s}^{-1}$, which gives a Jeans mass of about $10^5 \, M_\odot$, larger than the total gas mass in the box ($2 \times 10^4 \, M_\odot$). The injected turbulence was therefore efficient enough to prevent gas collapse; this was confirmed by the simulations where we have not observed any (complete) gas collapse.

**Table 1.** Common parameters to all simulations.

| Box size (pc) | BH initial velocity (km s$^{-1}$) | BH mass ($M_\odot$) | Gas density (cm$^{-3}$) |
|---|---|---|---|
| 40 | 15 | $10^4$ | 10 |

Small transient gas over-densities can appear, but they are rapidly dispersed by the turbulence.

About 10 Myr (corresponding to a turbulence box crossing time) are needed for the turbulence to settle and for the mean gas parameters, temperature, pressure, to stabilize. After this first period of 10 Myr, a BH was added, with a mass of $10^4\,M_\odot$ and an initial velocity of 15 km s$^{-1}$. Periodic boundary conditions allowed the BH (and the gas) to cross the box many times throughout one of our simulations. To avoid the BH crossing its own wake, the BH had an initial trajectory with an angle of 30° with respect to the $x$-axis and 60° with respect to the $z$-axis.

In this paper, we study the impact of resolution, turbulent wavenumber, and stochastic variation on the BH trajectory over time. The fixed parameters used for all simulations are summarised in Table 1, while parameter choices specific to a given simulation are reported in Table 2.

### 4.2. Parameter choices

#### 4.2.1. Run0

Run0 will serve as a reference case throughout this paper. It has been selected to have an intermediate level of turbulence, with $v_\mathrm{rms} = 3.2$ km s$^{-1}$, a wavenumber of $k = 8$, and $r_\mathrm{BHL}$ that is resolved throughout the simulation. As discussed in Sect. 4.1, the BHL radius increases with time as the BH velocity decreases. With a resolution of $\Delta x = 0.16$ pc for Run0, $r_\mathrm{BHL}$ evolves from $2.4\Delta x$ in the supersonic regime at the beginning of the simulation to $26\Delta x$ in the subsonic regime at the end of the simulation.

A projection of the gas density at time $t = 10$ Myr can be seen in the central panel of Fig. 2. For this simulation, typical eddy sizes are about 5 pc, which is slightly larger than $r_\mathrm{BHL}$, which ranges from 0.4 to 4 pc.

#### 4.2.2. Turbulent seed

Turbulence was initialised with a random seed, TSeed, which determines the exact instance of density peaks and troughs in the initial conditions of the simulation. For simulations with different TSeed, the gas has broadly the same appearance, with over- and under-densities having on average the same size and amplitude, but the local conditions encountered by the BH as it crosses the box differ. To test the statistical variation of results, we conducted a set of four simulations that had an identical setup to Run0 but were initialised with a different random seed. These simulations are called TSeedA, TSeedB, TSeedC, and TSeedD. The aim of this series of simulations is to estimate the effect of the stochastic processes at work and to assess the difference between the general trends of the gas and BH evolution and any random event that could happen.

#### 4.2.3. Resolution

To explore the response of the system and the dynamical friction force with respect to how well the BHL radius is resolved,

we had set up simulations for a range of resolutions, varying between $0.08 \leq \Delta x/\mathrm{pc} \leq 0.62$:
- with a $\Delta x = 0.62$ pc resolution (Res0.62pc), the BHL radius is unresolved ($r_\mathrm{BHL} \leq \Delta x$),
- with a $\Delta x = 0.31$ pc resolution (Res0.31pc), the BHL radius is marginally resolved ($r_\mathrm{BHL} \simeq \Delta x$) at the beginning of the run and resolved ($r_\mathrm{BHL} \geq 2.5\Delta x$) after 120 Myr,
- with a $\Delta x = 0.16$ pc resolution (Run0), the BHL radius is resolved,
- with a $\Delta x = 0.08$ pc resolution (Res0.08pc), the BHL radius is resolved, with $r_\mathrm{BHL}$ equal at least to $5\Delta x$.

#### 4.2.4. Turbulence wavenumber

To assess the impact of different sizes of turbulent eddies, we varied the turbulence wavenumber, that is the size of the individual turbulent regions as a function of the box size. Values tested here ranged from $k = 4$ to $k = 16$, leading to a driving scale of the turbulence from 10 pc to 2.5 pc:
- with a $k = 4$ turbulent wavenumber (k4, k4_TSeedA, k4_TSeedB simulations), the driving scale of the turbulence is about 10 pc, that is at least 3 times the BHL radius,
- with a $k = 8$ turbulent wavenumber (Run0, TSeedA, TSeedB, TSeedC, TSeedD, but also the simulations with various resolution Res0.08pc, Res0.31pc, Res0.62pc), the driving scale of the turbulence is about 5 pc, that is always larger than the BHL radius but ranging from 13 times larger at the beginning of the simulations to the same order of magnitude at the end,
- with a $k = 16$ turbulent wavenumber (k16, k16_TSeedA, k16_TSeedB simulations), the driving scale of the turbulence is about 2.5 pc, that is the order of magnitude of the BHL radius.

Snapshots of the resulting density distribution of simulations with $k = 4$ and 16 can be seen in Fig. 2 (respectively the left and right panels).

## 5. Results

### 5.1. Run0

#### 5.1.1. BH deceleration

When placed in a turbulent background medium, an over-dense wake develops downstream of a moving BH (see Fig. 3), which creates dynamical friction that is able to slow an originally supersonic BH (with initial Mach number $\mathcal{M}_\mathrm{ini} = 12.4$; this initial Mach number is derived from the average sound speed calculated from the simulation at the first output after BH insertion, at $t = 15$ Myr) down to the transonic regime (see Fig. 4, left panel) where the evolution stalls. This is consistent with expectations from analytical theory, where dynamical friction in the subsonic regime becomes negligibly efficient. We can note that the BH slows down to the sound speed, rather than to the turbulent (rms) velocity of the gas (here and after the rms velocity is measured for the gas outside a sphere of 5 pc radius centred on the BH to avoid including local increase in velocity caused by the growing overdensity of gas around the BH at late times in transonic regime). When the BH motion becomes transonic with respect to the turbulent Mach number (i.e. when $\mathcal{M}_{\rm turb} \simeq 1$), there is no noticeable effect.

**Table 2.** Specific simulation parameters.

| Name | Resolution (pc) | Ratio of BHL radius ($t = 0$) /resolution | Initial rms velocity (km s$^{-1}$) | Turbulent wavenumber | Turbulent seed |
|---|---|---|---|---|---|
| Run0 | 0.16 | 2.4 | 3.2 | 8 | 1 |
| Res0.08pc | 0.08 | 4.8 | 3.2 | 8 | 1 |
| Res0.31pc | 0.31 | 1.2 | 3.1 | 8 | 1 |
| Res0.62pc | 0.62 | 0.6 | 2.9 | 8 | 1 |
| TSeedA | 0.16 | 2.4 | 3.2 | 8 | 7 |
| TSeedB | 0.16 | 2.4 | 3.2 | 8 | 10 |
| TSeedC | 0.16 | 2.3 | 3.2 | 8 | 13 |
| TSeedD | 0.16 | 2.3 | 3.2 | 8 | 18 |
| k4 | 0.16 | 2.3 | 4.6 | 4 | 1 |
| k4_TSeedA | 0.16 | 2.3 | 4.6 | 4 | 6 |
| k4_TSeedB | 0.16 | 2.3 | 4.6 | 4 | 15 |
| k16 | 0.16 | 2.4 | 2.0 | 16 | 1 |
| k16_TSeedA | 0.16 | 2.4 | 2.0 | 16 | 9 |
| k16_TSeedB | 0.16 | 2.4 | 2.0 | 16 | 15 |

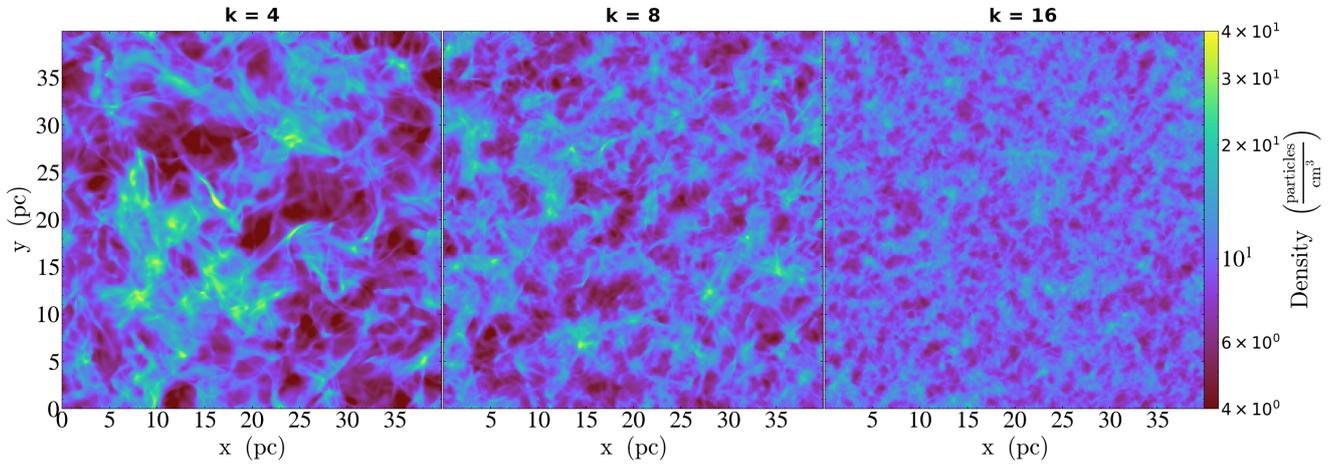

**Fig. 2.** Projection of the gas density. This at time $t = 10$ Myr, before the BH injection, for various turbulence wavenumbers: $k = 4$ (left panel, k4 simulation), $k = 8$ (middle panel, Run0 simulation), and $k = 16$ (right panel, k16 simulation).

During the simulation, the gaseous wake grows from a long and extended structure during the early supersonic phase to an almost spherical overdensity around $\mathcal{M} \approx 1$. Figure 3 shows the wake evolution after $t = 30, 85, 120$ Myr (supersonic regime), and 265 Myr (transonic regime) respectively.

The BH motion follows three main phases:
– Phase 1. During the first 110 Myr (from the BH injection at 10 Myr to about 120 Myr), the BH is already slowing down but at a low rate of about 0.04 km s$^{-1}$ Myr$^{-1}$ (Fig. 4, right panel, phase highlighted in red). During this phase, the BH-gas interaction is at work but appears to be limited and the dynamical friction not fully captured. This trend is confirmed with the resolution study in Sect. 5.3.
– Phase 2. During the next 100 Myr (from 120 Myr to 220 Myr), the decrease in BH velocity more than doubles in comparison to the first phase with a rate of 0.088 km s$^{-1}$ Myr$^{-1}$ (Fig. 4, right panel, yellow). The BH-gas interaction is reinforced and the dynamical friction is now well captured.
– Phase 3. After 220 Myr, the BH enters the transonic regime. During this phase, the BH velocity oscillates around the value of the sound speed but it does not decrease any more. At that time dynamical friction becomes inefficient and no further trends are observed until the end of the simulation. The BH velocity oscillates around 1.6 km s$^{-1}$ (Fig. 4, right panel, green), close to the gas sound speed of about 1.14 km s$^{-1}$ (while the rms velocity is larger at about 3 km s$^{-1}$).

We discuss the trends in detail in the parameter study sections, as trends can be quantified and explained more robustly when comparing Run0 to other simulations. Here we briefly note that the first two phases are related to how well the Bondi-Hoyle-Lyttleton radius is resolved. Early on, when $r_{\rm BHL}$ is barely resolved, with less than ~5 resolution elements, dynamical friction is not fully captured (first phase). Later on, $r_{\rm BHL}/\Delta x$ increases and dynamical friction becomes more efficient.

### 5.1.2. Development of the wake

In this section, we focus on the wake developing behind the BH, which is the over-dense gas responsible for the BH deceleration. In order to quantify the wake evolution, we measure the average gas density in a cylinder of radius 2 pc and length 20 pc just behind the BH, aligned with the BH velocity vector. The BH is injected at 10 Myr and the first available results shown here are at 15 Myr. The free fall time of gas in the BH potential for a BH of $10^4 \, M_\odot$ travelling at $15 \, \mathrm{km \, s^{-1}}$ in a gas with an effective velocity of $3 \, \mathrm{km \, s^{-1}}$ is about 0.04 Myr. By employing a sampling frequency of 5 Myr, we are therefore not following the set-up of the wake but its further development throughout the simulation. To have a reference value before the BH injection, we also measure the density in the same cylinder positioned according to the BH initial position and velocity direction in the first output of the simulation.

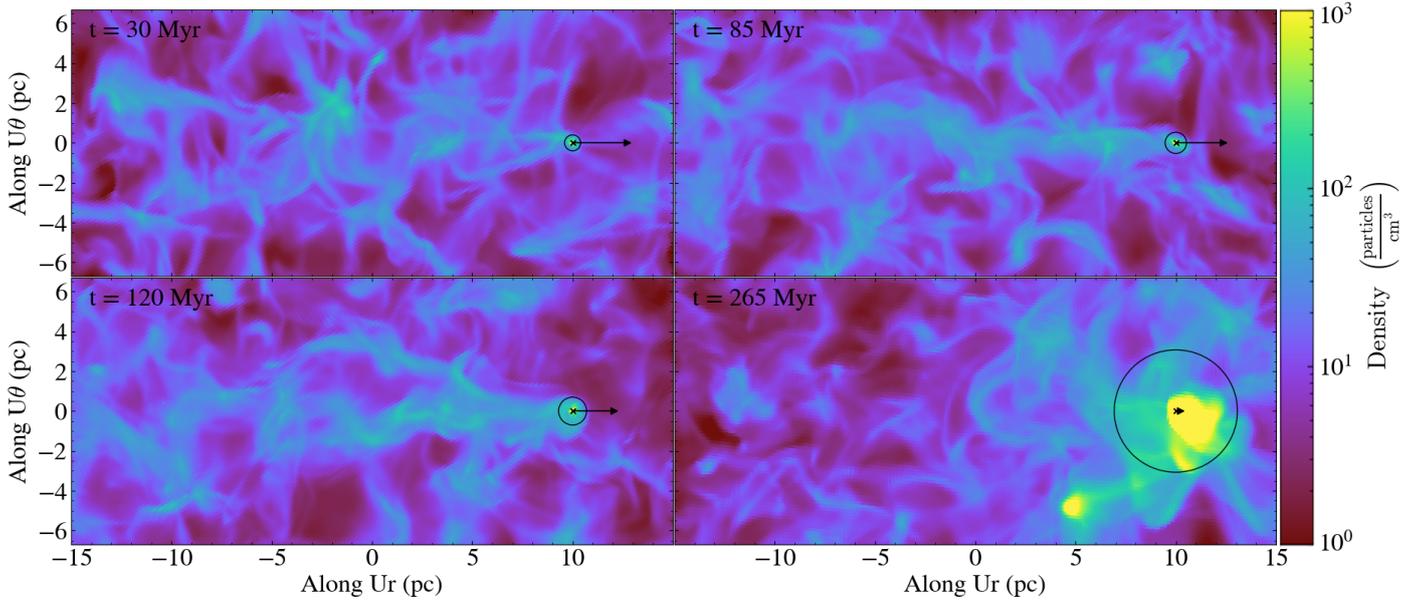

**Fig. 3.** BH wake evolution in Run0. The gas density is projected over a width of 5 pc at times $t = 30$ (top left), 85 (top right), and 120 (bottom left) Myr for the supersonic regime, and 265 Myr (bottom right) for the transonic regime. Images are displayed along the velocity vector of the BH (Ur). The cross marks the BH position, the circle shows the size of the BHL radius, and the length of the black arrow is proportional to the BH velocity. The images are oriented such that the BH velocity vector, and hence the gas wake as well, is contained in the plane of the image, with the gas wake on the left-hand side of the BH position.

Figure 5 shows the density evolution of the wake compared to the BH and gas velocity evolution for the Run0 simulation. The black dashed vertical lines are for the first BH output, 5 Myr after its injection ($t \sim 15$ Myr) and the last output in the supersonic regime respectively ($t \sim 220$ Myr). Before the BH injection, the density measured in the cylinder is about $10 \, \mathrm{cm^{-3}}$, in agreement with the gas mean density in the full box. After the BH injection the density quickly doubles to about $20 \, \mathrm{cm^{-3}}$ and continues to increase from there. The wake evolution is very different in the supersonic and in the transonic regimes: in the supersonic regime the wake density grows slowly at a rate of $2.6 \, \mathrm{cm^{-3}}$ per 10 Myr, whereas, in the transonic regime, the wake density shows large oscillations correlated with the BH relative velocity. Peaks in density correspond to low BH velocity and vice versa. Despite the oscillations, the average density of the wake continues to increase due to the gravity of the BH.

### 5.2. Sensitivity to initial turbulence seed

A turbulent gaseous medium is by definition a chaotic system, particularly sensitive to initial conditions. In order to assess the reliability and the stability of our results we run a set of simulations with the same physical and numerical conditions as Run0 but changing the turbulence seed. These are the simula-tions named TSeedA, TSeedB, TSeedC, and TSeedD for which the evolution of the BH relative velocity is presented in Fig. 6.

The results are quite homogeneous. TSeedA and TSeedB show a very similar evolution with a transition from phase 1 to phase 2 a little less pronounced than in Run0 but a transition to phase 3 at the same time (220 Myr). TSeedC and TSeedD on the other hand have a more noticeable transition between phase 1 and phase 2 leading to a slightly earlier transition to phase 3, about 10 to 20 Myr earlier (see vertical lines in Fig. 6). These differences are quite small and the global evolution of the BH due to dynamical friction is robust against small stochastic changes.

The consistency of our results was confirmed by repeating this analysis for simulations with other turbulence wavenumbers $k = 4$ and $k = 16$, which also showed good consistency. These results are available in Appendix A.

### 5.3. Importance of resolution

The resolution of our main simulation Run0 (0.16 pc) was chosen in order to resolve the BHL radius and indeed we were able to capture dynamical friction. To assess the effect of resolution, we here compare Run0 to one simulation with a higher resolution, one simulation with a lower resolution and one where the BHL radius is not resolved at any time. These are the simulations named Res0.08pc, Res0.31pc, and Res0.62pc respectively. The evolution of the BH relative velocity and the ratio between the BHL radius and resolution for this set of simulations is presented in the left and right panels of Fig. 7 respectively.

As can be seen in Fig. 7, the resolution has a significant impact on the results. When $\Delta x$ is comparable to or larger than the BHL radius, dynamical friction is not captured and the BH experiences little deceleration. This is the case for the Res0.62pc simulation, which has $r_\mathrm{BHL} \leq 1.2 \Delta x$ at all times, and is unable to build an overdensity within $r_\mathrm{BHL}$. And this is also the case for the first part of the Res0.31pc simulation, until 130 Myr, with a BHL radius lower than $1.5 \Delta x$. When $r_\mathrm{BHL} < 1.5 \Delta x$ (Fig. 7, right panel horizontal black dashed line) dynamical friction is not effective and the BH deceleration is minimal. The turning point at 130 Myr for the Res0.31pc simulation, when dynamical friction becomes efficient, can clearly be seen in Fig. 7 (left panel, yellow curve).

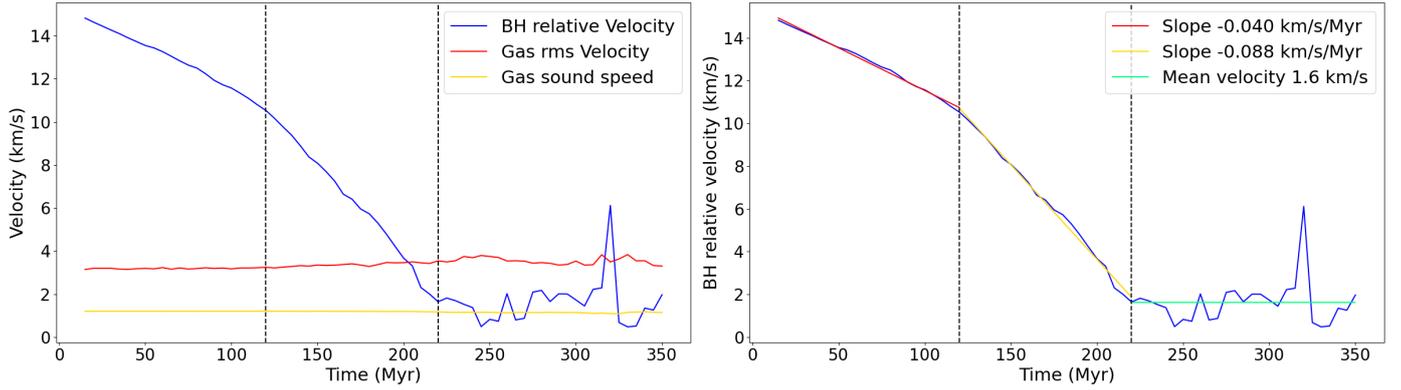

**Fig. 4.** Run0 velocities evolution. Left: evolution of the BH relative velocity (blue), the gas effective velocity (red, excluding a sphere of radius 5 pc around the BH), and sound speed (yellow) over time for Run0. Right: evolution of the BH relative velocity with superimposed fits to its time evolution. The dashed vertical lines in both panels mark the three phases of the BH deceleration. In the first phase the BHL radius is not well resolved, in the second the BHL radius is well resolved and deceleration is stronger, in the third the BH is in the transonic regime and does not decelerate further.

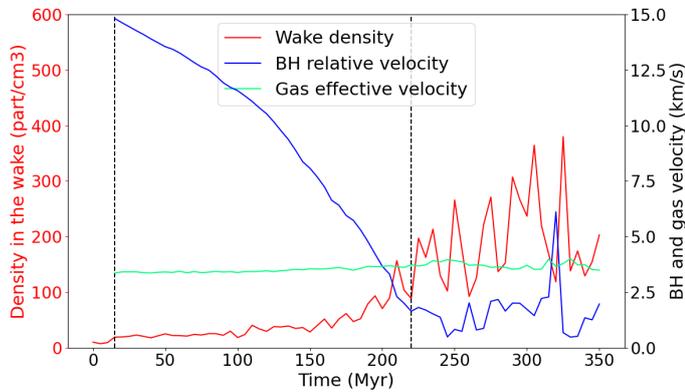

**Fig. 5.** Wake density evolution compared to the BH and gas velocity evolution for the Run0 simulation. The gas density is measured in a cylinder of 2 pc radius and 20 pc length just behind the BH, oriented along the BH velocity vector. In the measure of the gas effective velocity the gas within a sphere of radius 5 pc around the BH is excluded. The black dashed lines show the first BH output, 5 Myr after its injection and the last output in the supersonic regime respectively. In the supersonic regime density in the wake grows slowly and steadily, whereas, in the transonic regime, the density keeps increasing, but with large oscillations.

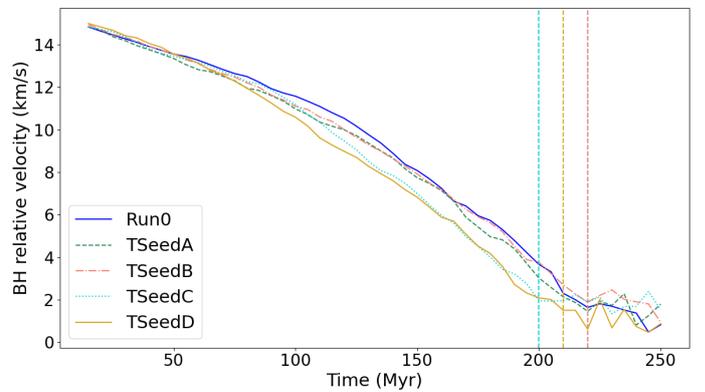

**Fig. 6.** Evolution with time of the BH relative velocity for different turbulence seed with a turbulence wavenumber $k = 8$. The vertical dashed lines mark the transition from super- to transonic; simulations Run0, TSeedA and TSeedB transition exactly at the same time, therefore only one line is visible (red dashed). This shows that our results are robust to small stochastic changes in the initial conditions.

When the BHL radius is larger than $1.5\Delta x$, dynamical friction is active but significantly lower than when $r_{BHL} > 4.5\Delta x$ (Fig. 7 right panel, horizontal black dotted line), at which point a (second) increase in slope can be seen in the evolution of velocities (vertical coloured dotted lines, left hand panel). We have already commented on this in Run0 with a change in slope of the BH velocity at 120 Myr (vertical red dotted line in Fig. 7), at which point the deceleration doubles to $0.088\,\mathrm{km\,s^{-1}\,Myr^{-1}}$. The same happens in the Res0.31pc simulation at 230 Myr (vertical yellow dotted line in Fig. 7) with the same deceleration rate around $0.088\,\mathrm{km\,s^{-1}\,Myr^{-1}}$ after this time. This result is again confirmed in Res0.08pc, which always has $r_{BHL} > 5\Delta x$, and where the deceleration rate is consistently $\sim 0.089\,\mathrm{km\,s^{-1}\,Myr^{-1}}$, except during the first 50 Myr when the wake is first forming. We therefore conclude that a minimum resolution of $r_{BHL}/\Delta x > 5$ is required to correctly capture the magnitude of dynamical friction in a turbulent medium, with results converged for the range of resolutions probed here out to $r_{BHL}/\Delta x \approx 50$.

By studying the evolution of wake density with simulation time for simulations with different resolutions, we can show that the wake grows faster with higher resolution in the supersonic regime (see Fig. 8). However, when comparing the wake evolution as a function of Mach number, rather than simulation time, a global pattern emerges. To compare the density across simulations in a time-independent manner, we highlight in Fig. 8 the time at which the BH first becomes trans-sonic using vertical dotted lines. Comparing the wake density at these points in time shows that density in the wake is similar for the three different resolution runs when the BH reaches $\mathcal{M} = 1$. In summary, in the supersonic regime, when increasing the resolution while keeping the same turbulence wavenumber, the density in the wake reaches the same level in a shorter time. In the transonic regime the mean peak density, ignoring the highest peaks and deepest troughs, appears to increase with resolution although in the Res0.31pc case we can not conclude whether the wake is still growing at the end of the simulation. The trend with resolution is consistent with gas being more easily bound to the BH

Lescaudron, S., et al.: A&A 674, A217 (2023)

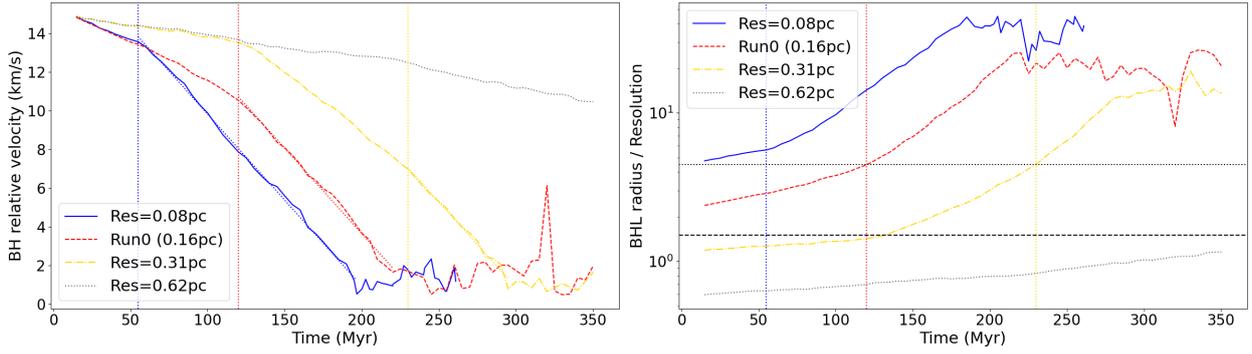

**Fig. 7.** Evolution with time of the BH relative velocity (left panel) and the BHL radius (right panel) for 4 different resolutions. The vertical lines show, for each simulation, the point where the BH deceleration reaches 0.09 km s$^{-1}$ Myr$^{-1}$ and the BHL radius about $4.5\Delta x$. The dashed and dotted horizontal lines in the right panel represent respectively BHL radius equal to $r_{BHL} = 1.5\Delta x$ and $r_{BHL} = 4.5\Delta x$. In the simulation where the BHL radius is unresolved (Res0.62pc), BH deceleration is very inefficient. In Res0.31pc, where the BHL radius is resolved after 120 Myr, we see at the same time that the BH starts to decelerate faster, with a second increase in deceleration occurring when $r_{BHL} = 4.5$ at $t = 240$ Myr. As the resolution increases and the BHL radius is resolved with more elements (Res0.16pc and Res0.08pc) the effect of dynamical friction becomes more pronounced.

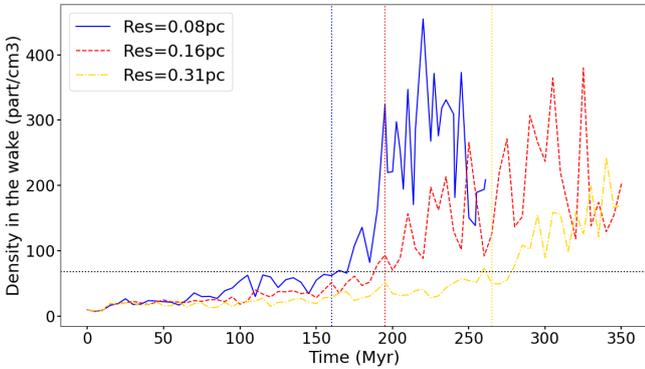

**Fig. 8.** Evolution of the density in the wake as a function of time for three different resolutions. The dotted vertical lines of the associated colours show the time of the transition from supersonic to subsonic regime in each case. At higher resolution the density in the wake increases faster, but stalls at similar values when the BH reaches $\mathcal{M} = 1$. The horizontal line shows the mean density in the wake at the time of the transition from supersonic to subsonic regime.

at high resolution. In the transonic regime, the wake becomes a more spherical overdensity, which continues to grow in mass, although it never reaches the state of a stable sphere.

### 5.4. Influence of turbulence wavenumber

Depending on the surrounding medium, BH may experience turbulence on different scales. We focus here on the influence of this turbulence scale on the dynamical friction efficiency. For high turbulence wavenumber $k$, the BH passes through many density peaks but each passage lasts a short time. With low $k$ the BH crosses a smaller number of bigger over-densities. The impact of changing wavenumber can be seen visually in Fig. 2.

Figure 9 shows the BH and gas rms velocity evolution for three different turbulence wavenumbers, with each wavenumber represented by three or four different turbulent seeds. As the simulations are done with a constant turbulence rms forcing, they have somewhat different initial rms velocities depending on the wavenumber: about 2 km s$^{-1}$ with $k = 16$, 3 km s$^{-1}$ with $k = 8$, and 4.5 km s$^{-1}$ with $k = 4$. In the calculation of the rms velocity a sphere of 5 pc radius has been excluded, in order to remove the contribution from gas infalling towards the black hole, which becomes large when the black hole reaches the transonic regime.

In the cases $k = 4$ and $k = 8$, the rms velocity remains roughly constant; whereas, in the case $k = 16$, it shows spikes up to 3.5 km s$^{-1}$. This is because the overdensity around the black hole becomes larger than 5 pc in this case, since the lesser effect of turbulence allows for the overdensity around the black hole to grow more easily.

The BH deceleration is slower in the $k = 4$ case than in the $k = 8$ and $k = 16$ cases because the wake is less stirred by the turbulence for high $k$, and can therefore persist for longer while growing somewhat more strongly. We speculate that this reduction in force for $k = 4$ occurs because a higher initial rms velocity makes the transfer of momentum and energy from the BH to the gas less efficient.

Performing the wake analysis on the simulations with different turbulence wavenumber, we show in Fig. 10 that the overdensity in the wake grows more strongly when the turbulence wavenumber is higher. The density reached by the wake at the end of the supersonic phase depends mainly on the turbulence wavenumber, and increases with it: from about $n = 35$ cm$^{-3}$ for $k = 4$, via 70 cm$^{-3}$ for $k = 8$ (Run0, Res0.08pc, and Res0.31pc) to about 150 cm$^{-3}$ for $k = 16$. This confirms that larger eddies are more efficient in dissipating the over-density behind the BH. In summary, in the supersonic regime, when increasing the turbulence wavenumber while keeping the same resolution, the density in the wake reaches a higher level, that is, it is stronger, over the same period of time.

Once the BH reaches the transonic regime, the wake density evolution is rather chaotic with strong oscillations, as can be seen in Fig. 10. In general, although the BH does not slow down any more, the wake density keeps increasing (except in the $k = 4$ case). This increase is actually stronger than in the supersonic regime (equal in the $k = 4$ cases) and is weaker with a lower turbulence wavenumber, since the growing overdensity around the BH is denser, larger, and more stable with higher $k$.

### 5.5. Comparison to analytic dynamical friction force

#### 5.5.1. Drag force as a function of Mach number

The last term in Eq. (3) includes $\ln(r_{max}/r_{min})$, which is sometimes referred to as the Coulomb logarithm. This term contains the only free parameters of the model. In this section, we compare the dynamical friction force experienced by the BH in our simulation to Eq. (3) for a range of different values of $\ln(r_{\max}/r_{\min})$. For easier comparison, we define the dimensionless drag force $f(\mathcal{M})$ as

$$F_{\rm DF}(\mathcal{M}) = f(\mathcal{M}) \frac{4\pi (GM_{\rm BH})^2 \rho_0}{c_s^2},$$

where $F_{\rm DF}$ is the dynamical friction force experienced by the BH. For the analytic values, $F_{\rm DF} = F_{\rm sup}$ from Eq. (3), so

$$f(\mathcal{M}) = \frac{1}{\mathcal{M}^2}\left[\frac{1}{2}\ln\left(1-\frac{1}{\mathcal{M}^2}\right) + \ln\frac{r_{\max}}{r_{\min}}\right]. \quad (6)$$

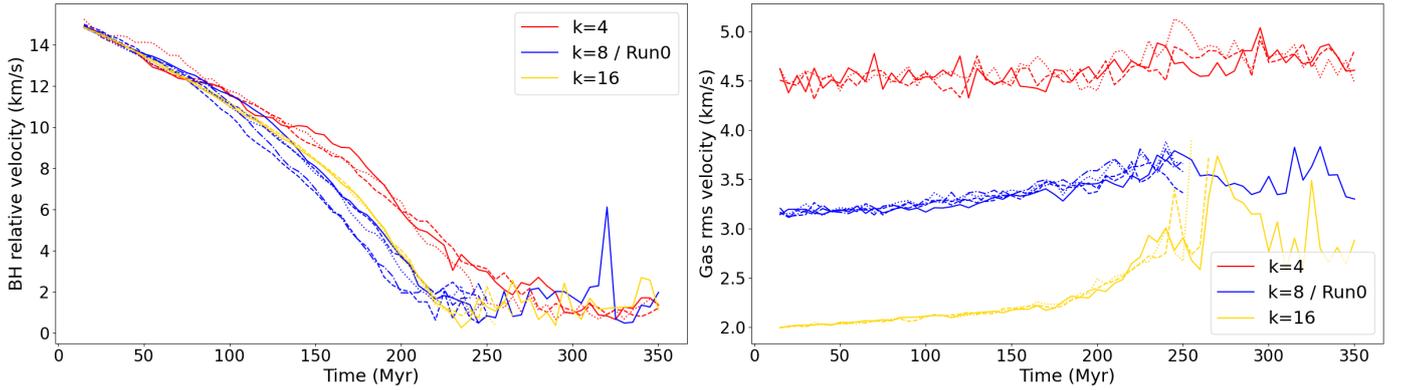

**Fig. 9.** Evolution with time of the BH relative velocity (left panel) and the gas rms velocity (right panel, excluding a sphere of radius 5 pc around the BH) for 3 different turbulent wavenumbers. Different linestyles of the same colour probe the stochastic variation of our results. Red curves show results for simulations k4, k4_TSeedA, and k4_TSeedB, blue curves show Run0, TSeedA, TSeedB, TSeedC, and TSeedD and yellow curves show k16, k16_TSeedA, and k16_TSeedB. Deceleration is slower at low wavenumbers, where, in our simulations, the rms velocity is higher, and plausibly transfer of kinetic energy from the BH to the gas less effective. Results are robust to stochastic variation of the turbulent density field.

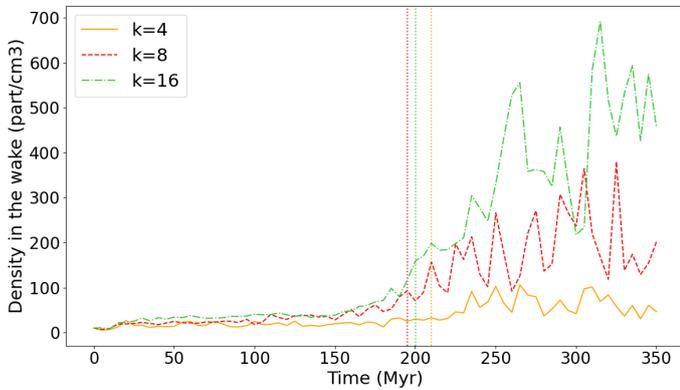

**Fig. 10.** Wake density as a function of time for turbulent wavenumbers $k = 4$, $k = 8$ (Run0), and $k = 16$. The dotted vertical lines mark the transition from supersonic to subsonic regime in each case. The density in the wake grows faster and to higher values at high wavenumbers, where the gas is closer to homogeneity. At low wavenumbers the larger eddies in the turbulence and the higher rms turbulence erase the wake more easily.

To calculate $F_{\rm DF}$ from our simulations, we compute the instantaneous acceleration of the BH by dividing its net velocity change during one simulation timestep by the length of that timestep.

The distribution of $f(\mathcal{M})$ as a function of Mach number can be seen in Fig. 11. To compare this to the analytic function, we plot curves computed using Eq. (3) for different values of $\ln(r_{\max}/r_{\min})$. Fitting the values of $\ln(r_{\max}/r_{\min})$ is not trivial because of the intrinsic scatter in the value of $f(\mathcal{M})$ introduced by the turbulence, so we prefer to bracket the distribution of data points from our simulation. We also add curves fitted on results for $\mathcal{M} > 4$, keeping in mind that these fits are quite poor.

Scatter in the measurement of $f(\mathcal{M})$ is introduced in two different ways. Firstly, the efficiency of the drag force depends on the background density, and therefore varies even at fixed Mach number as the BH traverses underdense or overdense regions. Secondly, the relative velocity between the BH and the background gas varies because the BH enters regions of higher or lower gas relative velocity, which in turn changes the Mach number we measure for our BH. This effects becomes stronger at lower Mach number, where velocity oscillations are more pronounced.

As can be seen in Fig. 11, the shape of $f(\mathcal{M})$ is significantly flattened in comparison to the analytic value at constant $\ln(r_{\max}/r_{\min})$. As a result, the value of $\ln(r_{\max}/r_{\min})$ that allows the analytic values from Ostriker (1999; Eq. (3)) to best match our simulations decreases significantly with Mach number. Phrased differently, if we were to fit $f(\mathcal{M})$ in a turbulent medium with Eq. (3) and a single $\ln(r_{\max}/r_{\min})$, the fit would under-predict $f(\mathcal{M})$ at high Mach number and over-predict it at low Mach number. We explore this phenomenon further in Sect. 5.5.2.

The turbulent wavenumber $k$ also affects the dispersion of the results: we observe a smoother evolution of $f(\mathcal{M})$ with high $k$ (see Fig. 11, bottom right panel for the $k = 16$ case) while the scatter increases with low $k$ (see Fig. 11, bottom left panel for the $k = 4$ case). This can be quantified using the values of $\ln(r_{\max}/r_{\min})$ which envelope all points, as plotted as dotted (minimum $\ln(r_{\max}/r_{\min})$) and dashed (maximum $\ln(r_{\max}/r_{\min})$) lines in Fig. 11. For $k = 4$, $\ln(r_{\max}/r_{\min})$ ranges from 0 to 15, while the range tightens to 0.05 to 12 in the $k = 8$ case, and even further to 0.1 to 8 in the $k = 16$ case. Also shown on the figure are fitted curves for all values of $f(\mathcal{M})$ for $\mathcal{M} > 4$ (dashed-dotted lines). These fits do not reveal any strong dependence on $k$ as the fit is generally poor due to the stochastic nature of the problem and the decrease in efficiency of dynamical friction with decreasing Mach number in our simulation.

When looking at the impact of resolution (Fig. 11, top left panel), dynamical friction becomes more efficient at higher resolution, as can be seen by the fact that the values of $f(\mathcal{M})$ for $\Delta x = 0.31$ pc (red) generally lie below those for $\Delta x = 0.08$ pc (green). However, higher resolution simulations also show more scatter, and therefore require a broader range of $\ln(r_{\max}/r_{\min})$ to envelop the data ($0.05 < \ln(r_{max}/r_{min}) < 17$ if $\Delta x = 0.08$ pc, compared to $0.05 < \ln(r_{max}/r_{min}) < 10$ if $\Delta x = 0.31$ pc).

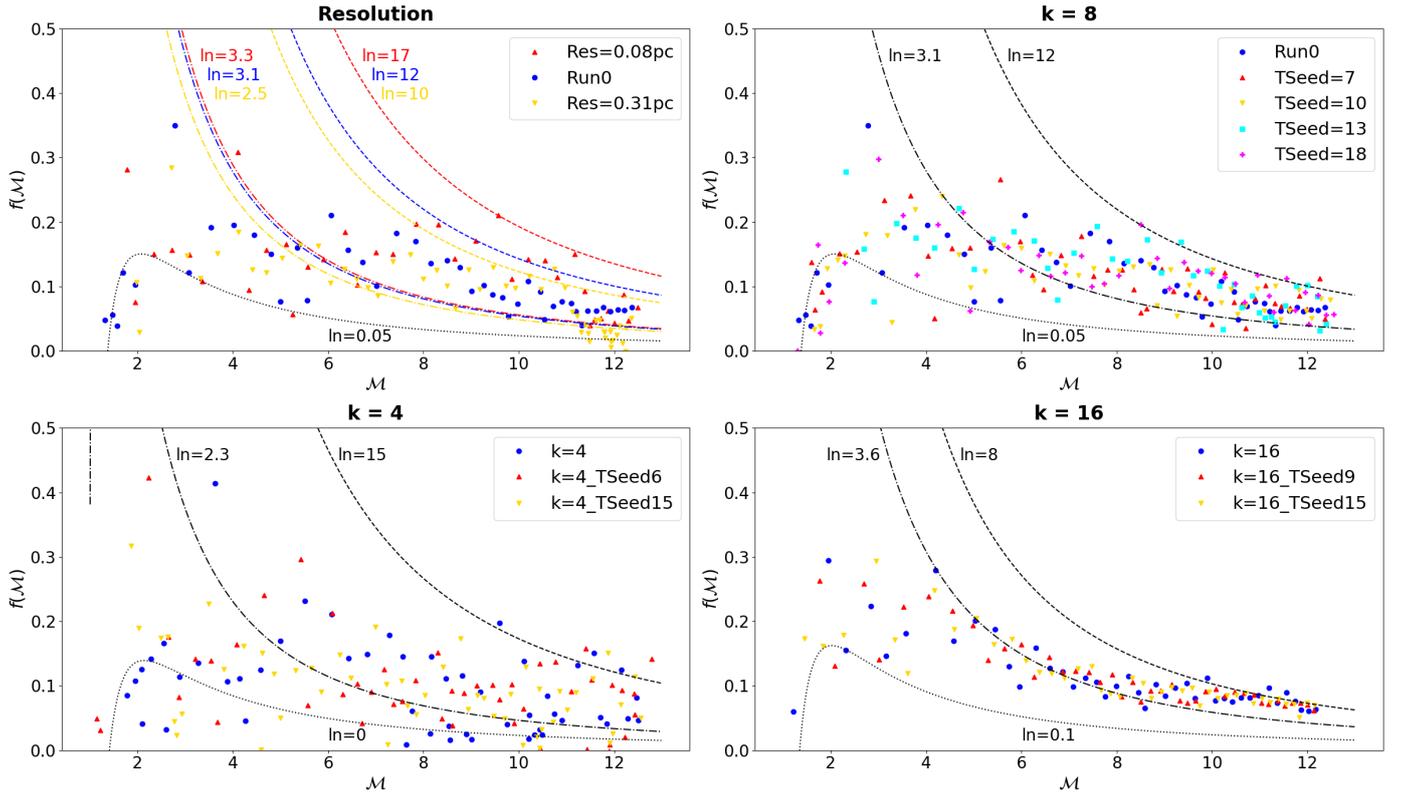

**Fig. 11.** Magnitude of the dimensionless drag force $f(\mathcal{M})$ as defined in Eq. (6) is shown as a function of Mach number $\mathcal{M}$. Top left: resolution study for $k = 8$. Top right to bottom left (clockwise): simulations with $k = 8$, $k = 16$, and $k = 4$ respectively, all with resolution 0.16 pc and different turbulent seeds shown as different colours. The curves represent Eq. (6) for different values of $\ln(r_{max}/r_{min})$: the dotted curves envelope the lower points with a minimum $\ln(r_{max}/r_{min})$, the dashed curves envelope the upper points with a maximum $\ln(r_{max}/r_{min})$, and the dash-dotted lines are a fit for $\mathcal{M} > 4$. The corresponding values of $\ln(r_{max}/r_{min})$ are given on the plots. The main impact of $\mathcal{M}$ on $f(\mathcal{M})$ is not a variation in the magnitude of $f(\mathcal{M})$ but an increase in its scatter for decreasing $\mathcal{M}$. At high wavenumber, when we are closer to the homogeneous case, the magnitude of $f(\mathcal{M})$ varies more with $\mathcal{M}$, while in the least homogeneous case ($k = 4$) scatter dominates at all $\mathcal{M}$.

In conclusion, the dynamical friction forces as a function of Mach number on the BH in a turbulent medium have an intrinsically flatter shape than in the homogeneous case derived in Ostriker (1999). In a turbulent medium, at fixed $\mathcal{M}$, the magnitude of the force increases for higher resolution and higher $k$.

### 5.5.2. Efficiency of the drag force in a turbulent medium

Given that our simulation has a uniform resolution, $r_{min} = \Delta x$ is well defined for our simulations. This leaves $r_{max}$ as the only free parameter in $\ln(r_{max}/r_{min})$ to quantify the magnitude of the dynamical friction. We speculated as to the expected values of $r_{max}$ for a turbulent medium in Sect. 2. In this section, we fit the magnitude of the drag force measured from the simulation using Eq. (3) to measure $r_{max}$, assuming $r_{min} = \Delta x$. We note that $r_{max}$ discussed here is not the physical extent of the wake behind the BH, which we shall refer to as $r_{wake}$ instead, but reflects the net deceleration of the BH from both the wake and any other gravitational forces on the BH due to turbulent over-densities.

The results for this fit as a function of Mach number can be seen in Fig. 12. Two things are immediately apparent: $r_{max}$ is strongly a function of Mach number, and the values found here range from unphysically small ($r_{max} \leq r_{min}$) to unreasonably large ($r_{max} \gg l_{box}$). The physical extent of the wake, $r_{wake}$ is difficult to measure in simulations due to the density structures inherent in turbulent gas, which make it difficult to differentiate between the overdensity caused by the BH and an overdensity that is simply part of the turbulent gas. However, it certainly cannot exceed the size of the box. This sets a physical upper limit on $r_{wake}$ of around 40 pc. The $r_{max}$ required to explain the force according to Ostriker (1999) is also clearly higher than the theoretical estimate for $r_{wake}$ in Eq. (5), which remains 2–3 pc throughout the simulation.

The magnitude of $r_{max}$ is a strongly increasing function of Mach number, which reflects the flattened distribution of $f(\mathcal{M})$ as a function of Mach number in comparison to the homogeneous case (see Sect. 5.5.1 for a discussion). This is in clear contradiction with both the original prediction by Ostriker (1999), who predict that $r_{max} \propto \mathcal{M}t$, and the trend predicted in Sect. (2) where $r_{max} \propto \mathcal{M}^{-1}$ (see Eq. (5)). Extrapolating from our turbulent simulations to a smooth case (i.e. one for which $k \to \infty$) is not trivial, as discussed in Appendix B. Nevertheless, using a simulation as close to Run0 as feasible, we confirmed that the $k = 16$ case presented here shows good convergence with a fully smooth case for $\mathcal{M} > 10$, while at lower Mach numbers the turbulent case is always below the smooth run, highlighting that the suppression of the force at low Mach numbers is caused by turbulence (see Appendix B for details). Another value for comparison for a locally smooth medium can be taken from Chapon et al. (2011), who used the dynamical evolution of a transonic BH in a smooth galaxy in the range $\mathcal{M} = 1.3$ to 0.8 to predict $r_{max}/r_{min} \sim 25$, that is $r_{max} \approx 4$ pc for our resolution.

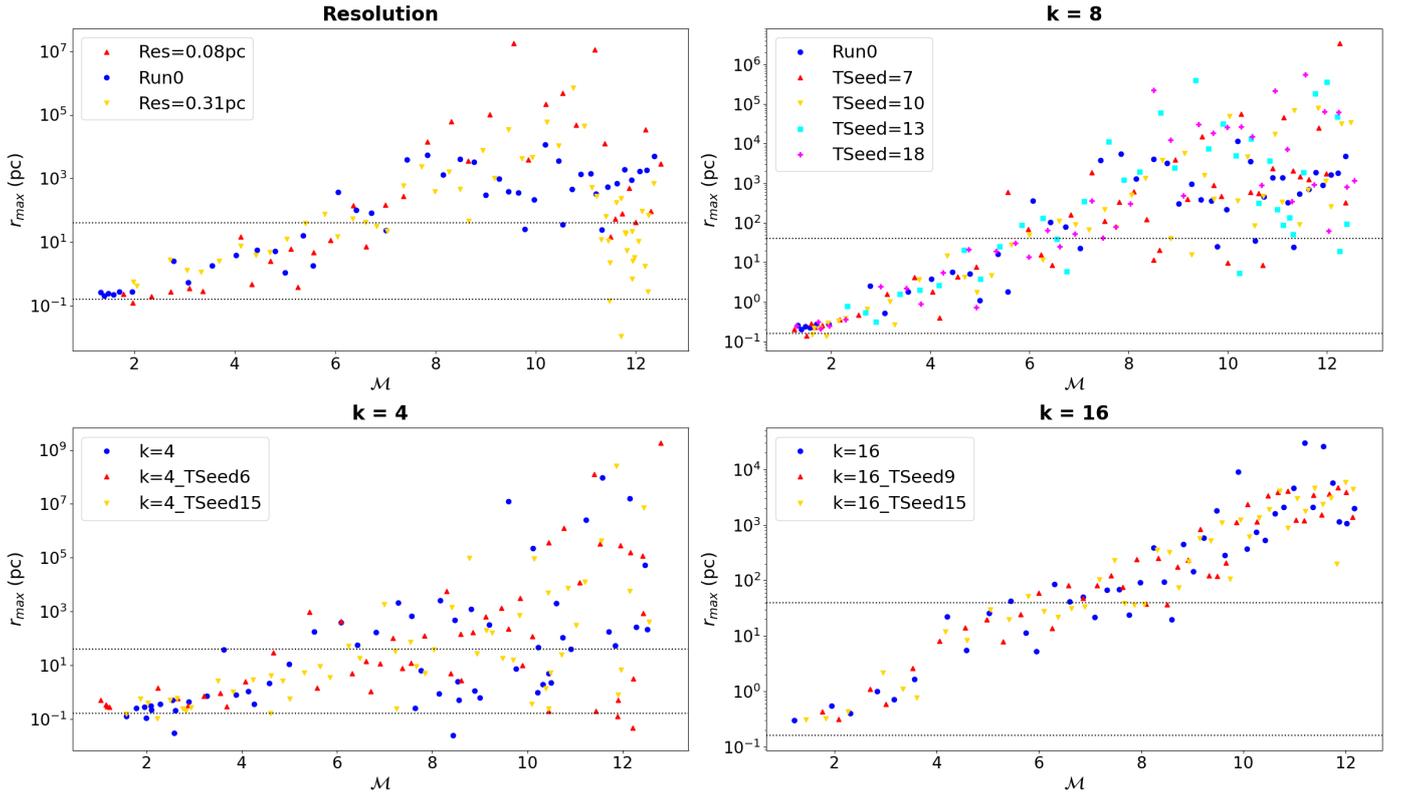

**Fig. 12.** $r_{max}$, calculated from Eq. (5), as a function of Mach number $\mathcal{M}$. Top left: resolution study for $k = 8$. Top right to bottom left (clockwise): simulations with $k = 8$, $k = 16$, and $k = 4$ respectively, all with resolution 0.16 pc and different turbulent seeds shown as different colours. The two horizontal dotted lines indicates the size of the box (40 pc) and the fiducial resolution (0.16 pc). The values obtained for high $\mathcal{M}$ are unreasonably high, meaning that the force experienced by a BH in a turbulent medium is not well approximated by the analytic formalism developed for a homogeneous medium.

This is in good agreement with our results for both high turbulent Mach numbers (see Fig. 12) and the smooth simulations (see Appendix B).

We conclude that the net long-term dynamical friction in a turbulent medium, once fully resolved, is somewhat more efficient than in the analytical case for strongly supersonic BHs (of $\mathcal{M} > 5$) but significantly less efficient as the BH approaches the transonic regime. Finally, we note that the difference in $r_{max}/r_{min}$ is exacerbated by the exponential in the functional form, and also that in the scale-free formalism of Ostriker (1999) $r_{min}$ is not fully defined, while in simulations $r_{min}$ is necessarily defined to be the resolution. Compared instead to a homogeneous case simulation as described in the Appendix B, dynamical friction in a turbulent medium, once fully resolved, is similar at high Mach numbers and less efficient close to and at the transonic regime.

As can be seen in Fig. 12 (top right and bottom panels), lower turbulent wavenumbers produce significantly more scatter in $r_{max}$ but no significant difference in the average $r_{max}$ at a given Mach number. This is in good agreement with the fact that the BH velocity evolution in Fig. 9 shows a small delay in deceleration for $k = 4$ and little difference between $k = 8$ and $k = 16$.

One caveat to this conclusion is that the magnitude of $r_{max}$ calculated using Eq. (3) is very sensitive to the values of the constants, such as the BH mass. $\rho_0$ is particularly difficult to define. Here it is taken to be the average background density, but obviously the BH passes a series of over and under-densities so $\rho_0$ is poorly defined in this context. Another hint that a turbulent medium cannot be well approximated by $\rho_0$ comes from the fact that there is a clear dependence of the maximum $r_{max}$ on the turbulent wavenumber $k$: higher wavenumbers lead to lower maximum values of $r_{max}$ and less scatter (Fig. 12, bottom panels). This suggests that $v_{rms}$ also plays a role in determining the magnitude of the force. Overall, we conclude that dynamical friction due to a turbulent medium for strongly supersonic BH is more efficient than in the homogeneous case, but less efficient as the BH approaches the transonic regime.

## 6. Conclusions

In this paper, we presented a study of dynamical friction in a turbulent gaseous medium. We investigated how the deceleration of an initially supersonic BH depends on the properties of the local turbulent medium, thereby bridging the gap between isolated numerical experiments and full-scale galaxy evolution simulations. We find that:

- In a turbulent medium, an over-dense wake develops downstream of the BH, which slows an originally supersonic BH (here, $\mathcal{M}_{ini} \sim 10$) down to the transonic regime, where $\mathcal{M} = 1$.
- The efficiency of dynamical friction depends on how well the Bondi-Hoyle-Lyttleton radius is resolved in the simulations. For $1 < r_{BHL}/\Delta x < 5$, the dynamical friction is present but reduced in comparison to higher values. At higher resolution, the magnitude of the force is converged for the full range of resolutions tested here, $5 < r_{BHL}/\Delta x < 50$.
- Even in a turbulent medium, evolution of dynamical friction is determined by the classic thermal Mach number $\mathcal{M} = v_{rel}/c_s$ rather than the turbulent Mach number $\mathcal{M}_{turb} = v_{rel}/\sqrt{c_s^2 + v_{rms}^2}$ which takes the turbulent rms velocity into account. While nothing special happens when $\mathcal{M}_{turb} = 1$, dynamical friction becomes very inefficient when $\mathcal{M} = 1$ and the BH velocity evolution stalls at this point with no further slow-down observed in our simulations.
- Dynamical friction is more efficient for higher wavenumbers of the background turbulence. In our simulations, the BH reaches the transonic regime after approximately 200 Myr for $k = 16$, rather than the 300 Myr required when $k = 4$. This result is robust to the statistical variations that occur when drawing a different instance of the same turbulent density distribution.
- In comparison to dynamical friction in a homogeneous medium, presented in Ostriker (1999), we report that for a decelerating BH in a turbulent medium, magnitude of the force depends only weakly on $\mathcal{M}$ as long as $\mathcal{M} > 1$. Overall, the $r_{max}$ required to fit the formula from Ostriker (1999) is unrealistically high for high $\mathcal{M}$, where we find that $r_{max}$ exceeds the size of the box probed here by several orders of magnitude. By contrast, $r_{max} < \Delta x$ for low $\mathcal{M}$ at low values of turbulent wavenumber $k$, but approaches physically reasonable values of a few pc as $k \to \infty$. We therefore report that in a turbulent medium, dynamical friction is more efficient for high Mach numbers, but less efficient as the BH approaches the transonic regime (i.e. as $\mathcal{M} \to 1$).
- At a given Mach number, scatter in dynamical friction magnitude is larger for turbulence with lower $k$ than for higher $k$, due to the more extended but less frequent density perturbations traversed by the BH.

There are a few caveats to the work presented here. One is that we do not capture the exchange of momentum between the BH and the background gas due to accretion. Using the BHL accretion rate

$$\dot{M}_{BH} = \frac{4\pi\rho G^2 M_{BH}^2}{(v_{rel}^2 + v_{eff}^2)^{3/2}},$$

we estimate that the mass accretion rate during the supersonic regime with $v_{rel} = 10\,\text{km}\,\text{s}^{-1}$, $v_{rms} = 4\,\text{km}\,\text{s}^{-1}$ is $\sim 5\,M_\odot\,\text{Myr}^{-1}$. This leads to about $10^3\,M_\odot$ accreted during the 200 Myr of supersonic regime, and, hence, a decrease of approximately a tenth of the initial BH velocity due to gas accretion. This effect could potentially allow the BH to slow down into the sub-sonic regime, as accretion becomes more efficient at low Mach numbers. Including the impact of accretion in the long-term evolution of the BH is left to future work.

On the other hand, the aftermath of gas accretion, involving injection of energy as feedback, has been shown to act counter to dynamical friction. Indeed, with radiative (Park & Bogdanović 2017), thermal (Souza Lima et al. 2017), and kinetic (Gruzinov et al. 2020) feedback the wake can be destroyed, nullifying the deceleration. Furthermore, feedback can create an underdense region behind the BH and the net effect is for the BH to experience an acceleration in the direction of motion, at least temporarily and for some gas densities and BH masses (Toyouchi et al. 2020). We caution that the small size of our box means that the BH traverses the same region of gas repeatedly throughout the study, despite choosing an angle that minimises this effect. The turbulent forcing implemented here re-randomises the gas density field before the next pass, but we cannot rule out that cumulative effects might affect the late-time evolution of our BH.

Overall, we conclude that dynamical friction due to a turbulent background medium is able to efficiently slow a BH down to $\mathcal{M} = 1$, but cannot further decelerate the BH. This force is self-consistently captured by hydrodynamical simulations as long as the minimum resolution criterion of $r_{BH}/\Delta x > 5$ is met, but under-resolved otherwise. When resolved, the force in a turbulent medium exceeds that in a homogeneous medium at high Mach number, but is reduced around the trans-sonic regime.

*Acknowledgements.* R.S.B. gratefully acknowledges funding from Newnham College, University of Cambridge and the ANR grant LYRICS (ANR-16-CE31-0011). This work was granted access to the HPC resources of CINES under the allocations A0080406955 and A0100406955 made by GENCI. This work has made use of the Infinity Cluster hosted by Institut d'Astrophysique de Paris. We thank Stéphane Rouberol for smoothly running this cluster for us. Visualisations in this paper were produced using the YT PROJECT (Turk et al. 2011).

## References

Amaro-Seoane, P., Audley, H., Babak, S., et al. 2017, ArXiv e-prints [arXiv:1702.00786]
Bahé, Y. M., Schaye, J., Schaller, M., et al. 2022, MNRAS, 516, 167
Barausse, E., Dvorkin, I., Tremmel, M., Volonteri, M., & Bonetti, M. 2020, ApJ, 904, 16
Bartlett, D. J., Desmond, H., Devriendt, J., Ferreira, P. G., & Slyz, A. 2021, MNRAS, 500, 4639
Beattie, J. R., & Federrath, C. 2020, MNRAS, 492, 668
Beckmann, R. S., Devriendt, J., Slyz, A., et al. 2017, MNRAS, 472, 949
Beckmann, R. S., Slyz, A., & Devriendt, J. 2018, MNRAS, 478, 995
Bellovary, J. M., Governato, F., Quinn, T. R., et al. 2010, ApJ, 721, L148
Bellovary, J. M., Cleary, C. E., Munshi, F., et al. 2019, MNRAS, 482, 2913
Bellovary, J. M., Hayoune, S., Chafla, K., et al. 2021, MNRAS, 505, 5129
Blecha, L., Cox, T. J., Loeb, A., & Hernquist, L. 2011, MNRAS, 412, 2154
Bleuler, A., & Teyssier, R. 2014, MNRAS, 445, 4015
Bower, R. G., Benson, A. J., Malbon, R., et al. 2006, MNRAS, 370, 645
Chandrasekhar, S. 1943, ApJ, 97, 255
Chapon, D., Mayer, L., & Teyssier, R. 2011, MNRAS, 429, 3114
Chen, N., Ni, Y., Tremmel, M., et al. 2022, MNRAS, 510, 531
Cho, J., & Vishniac, E. T. 2000, ApJ, 539, 273
Colpi, M., Mayer, L., & Governato, F. 1999, ApJ, 525, 720
Commerçon, B., Marcowith, A., & Dubois, Y. 2019, A&A, 622, A143
Cowie, L. L. 1977, MNRAS, 180, 491
Croton, D. J., Springel, V., White, S. D. M., et al. 2006, MNRAS, 365, 11
Di Matteo, T., Springel, V., & Hernquist, L. 2005, Nature, 433, 604
Dokuchaev, V. P. 1964, Soviet Astron., 8, 23
Dosopoulou, F., & Antonini, F. 2017, ApJ, 840, 31
Dotti, M., Colpi, M., Haardt, F., & Mayer, L. 2007, MNRAS, 379, 956
Dubois, Y., Devriendt, J., Slyz, A., & Teyssier, R. 2010, MNRAS, 409, 985
Dubois, Y., Devriendt, J., Slyz, A., & Teyssier, R. 2012, MNRAS, 420, 2662
Dubois, Y., Pichon, C., Devriendt, J., et al. 2013, MNRAS, 428, 2885
Dubois, Y., Peirani, S., Pichon, C., et al. 2016, MNRAS, 463, 3948
Dubois, Y., Beckmann, R., Bournaud, F., et al. 2021, A&A, 651, A109
Eswaran, V., & Pope, S. B. 1988, Comput. Fluids, 16, 257
Evans, C. R., & Hawley, J. F. 1988, ApJ, 332, 659
Federrath, C., Roman-Duval, J., Klessen, R. S., Schmidt, W., & Mac Low, M. M. 2010, A&A, 512, A81
Foglizzo, T., Galletti, P., & Ruffert, M. 2005, A&A, 435, 397
Fromang, S., Hennebelle, P., & Teyssier, R. 2006, A&A, 457, 371
Fujii, M., Funato, Y., & Makino, J. 2006, PASJ, 58, 743
Goldreich, P., & Sridhar, S. 1995, ApJ, 438, 763
Granato, G. L., De Zotti, G., Silva, L., Bressan, A., & Danese, L. 2004, ApJ, 600, 580
Gruzinov, A., Levin, Y., & Matzner, C. D. 2020, MNRAS, 492, 2755
Guillet, T., & Teyssier, R. 2011, J. Comput. Phys., 230, 4756
Hurley, J. R., Tout, C. A., & Pols, O. R. 2002, MNRAS, 329, 897
Iben, I., Jr., & Livio, M. 1993, PASP, 105, 1373
Jenet, F. A., Lommen, A., Larson, S. L., & Wen, L. 2004, ApJ, 606, 799
Jenet, F. A., Hobbs, G. B., Lee, K. J., & Manchester, R. N. 2005, ApJ, 625, L123
Korol, V., Ciotti, L., & Pellegrini, S. 2016, MNRAS, 460, 1188
Kunyang, Li, Bogdanović, T., Ballantyne, D. R., & Bonetti, M. 2022, ApJ, 933, 104
Lapiner, S., Dekel, A., & Dubois, Y. 2021, MNRAS, 505, 172
Li, K., Bogdanović, T., & Ballantyne, D. R. 2020a, ApJ, 896, 113
Li, K., Bogdanović, T., & Ballantyne, D. R. 2020b, ApJ, 905, 123
Ma, L., Hopkins, P. F., Ma, X., et al. 2021, MNRAS, 508, 1973
MacLeod, M., & Ramirez-Ruiz, E. 2015, ApJ, 803, 41
Mayer, L. 2013, Class. Quant. Grav., 30, 244008
Miyoshi, T., & Kusano, K. 2005, J. Comput. Phys., 208, 315

Morton, B., Khochfar, S., & Oñorbe, J. 2021, MNRAS, submitted [arXiv:2103.15848]
Ni, Y., Di Matteo, T., Bird, S., et al. 2022, MNRAS, 513, 670
Ogiya, G., & Burkert, A. 2016, MNRAS, 457, 2164
Ostriker, E. C. 1999, ApJ, 513, 252
Park, K., & Bogdanović, T. 2017, ApJ, 838, 103
Park, K., & Bogdanović, T. 2019, ApJ, 883, 209
Pfister, H., Volonteri, M., Dubois, Y., Dotti, M., & Colpi, M. 2019, MNRAS, 486, 101
Reines, A. E., Condon, J. J., Darling, J., & Greene, J. E. 2020, ApJ, 888, 36
Rephaeli, Y., & Salpeter, E. E. 1980, ApJ, 240, 20
Ricarte, A., Tremmel, M., Natarajan, P., Zimmer, C., & Quinn, T. 2021, MNRAS, 503, 6098
Ruderman, M. A., & Spiegel, E. A. 1971, ApJ, 165, 1
Ruffert, M. 1996, A&A, 311, 817
Ruffert, M., & Arnett, D. 1994, ApJ, 427, 351
Schmidt, W., Federrath, C., Hupp, M., Kern, S., & Niemeyer, J. C. 2009, A&A, 494, 127
Sijacki, D., Springel, V., Di Matteo, T., & Hernquist, L. 2007, MNRAS, 380, 877
Souza Lima, R., Mayer, L., Capelo, P. R., & Bellovary, J. M. 2017, ApJ, 838, 13
Sutherland, R. S., & Dopita, M. A. 1993, ApJS, 88, 253
Teyssier, R. 2002, A&A, 385, 337
Teyssier, R., Fromang, S., & Dormy, E. 2006, J. Comput. Phys., 218, 44
Toyouchi, D., Hosokawa, T., Sugimura, K., & Kuiper, R. 2020, MNRAS, 496, 1909
Tremmel, M., Governato, F., Volonteri, M., & Quinn, T. R. 2015, MNRAS, 451, 1868
Tremmel, M., Karcher, M., Governato, F., et al. 2017, MNRAS, 470, 1121
Turk, M. J., Smith, B. D., Oishi, J. S., et al. 2011, ApJS, 192, 9
Volonteri, M., & Perna, R. 2005, MNRAS, 358, 913
Volonteri, M., Pfister, H., Beckmann, R. S., et al. 2020, MNRAS, 498, 2219
Yu, Q. 2002, MNRAS, 331, 935
Zhu, H., & Gnedin, N. Y. 2021, ApJS, 254, 12

## Appendix A: Dependence on random seed

In Figs. A.1 and A.2 we show the analogues of Fig. 6 for $k = 4$ and $k = 16$. As for $k = 8$ the results are robust to small stochastic changes in the initial conditions.

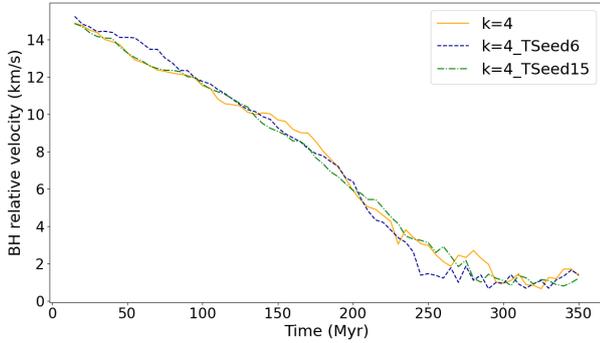

**Fig. A.1.** Evolution with time of the BH relative velocity for different turbulence seed with a turbulence wavenumber $k = 4$.

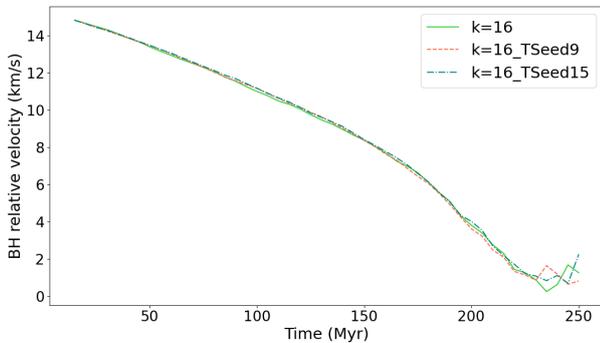

**Fig. A.2.** Evolution with time of the BH relative velocity for different turbulence seed with a turbulence wavenumber $k = 16$.

## Appendix B: Dynamical friction in a smooth medium

Creating a smooth equivalent to our turbulent boxes is not trivial for several reasons. Firstly, without the reheating from turbulence, the gas in the box would bulk-cool over the timescales required for the dynamic study. When the gas is not allowed to cool, it is instead dynamically heated by the repeated passage of the BH, and also picks up a non-negligible bulk velocity through momentum transfer from the BH. Both effects artificially speed up the decay of the Mach number, invalidating measurements of the drag force as a function of Mach number. One solution is to use a bigger simulation box, as this increases the thermal and dynamical mass of the gas, but maintaining constant resolution for bigger boxes quickly becomes numerically unfeasible. Through experiment, we found that doubling the box size of our fiducial simulation, Run0, allowed for a reasonable compromise: for an 80 pc box, a simulation with the same input parameters as Run0 (including resolution) sees a final sound speed increased by a factor ~1.5 and a final bulk speed of the gas of only 0.6 km/s. There is also the problem that the gas does not stay smooth over the timescales required for our setup: old wakes from the repeated passage of the BH through the box create density fluctuations encountered by the BH during its next passage. To tackle these issue, we use a set of simulations that have the same parameters as Run0, but have initially uniform density (and no turbulent forcing), an 80 pc box and a starting Mach numbers of $\mathcal{M}_{ini}$ = 12.5, 10, 7.5, 5, and 2.5 respectively. We evolve each simulation only for $\Delta \mathcal{M} = 3$, that is for a limited change in Mach number.

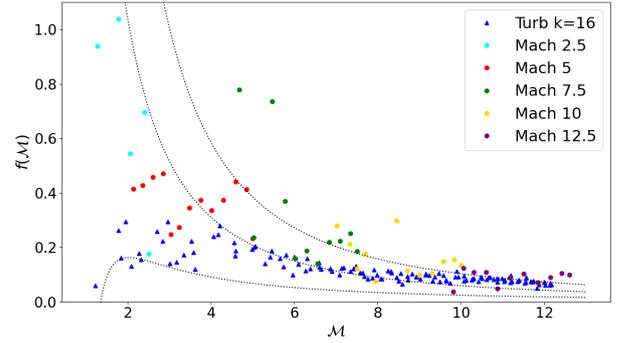

**Fig. B.1.** Dimensionless drag force $f$ on the black hole for a suite of non-turbulent equivalents of Run0 with different initial Mach numbers, and the points of our $k = 16$ simulation for comparison. Black lines represent the analytic force according to Ostriker (1999) for the same values of Coulomb logarithm than in the Fig. 11 for the $k = 16$ case: 0.1, 3.6, 8.

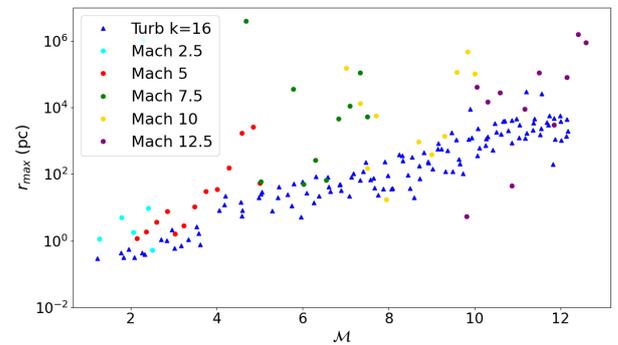

**Fig. B.2.** $r_{max}$ (calculated using Eq. 5) in the paper for non-turbulent versions of Run0 with different initial Mach numbers. Our turbulent $k = 16$ simulation is for comparison.

Comparing the dimensionless drag force on the BH for our non-turbulent boxes (Fig. B.1) to the highest wavenumber from the turbulent set (bottom right panel of Fig. 11), the evolution of the force as a function of the Mach number in a smooth medium is similar to highest wave-number case, $k = 16$, but it has a higher magnitude at low $\mathcal{M}$. This is consistent with the previous trend of higher forces at low Mach number for increasing $k$, and suggests that the suppression of the drag force in the transonic regime is due to the turbulence. The same is true for $r_{max}$ shown in Fig. B.2, which show large values at high Mach numbers and reasonable results of a few pc in the trans-sonic regime. $r_{max} = 5$ would mean $\ln(\lambda) = 3.44$, in reasonable agreement with both Chapon et al. (2011; $\ln(\Lambda) = 3.2$) and Beckmann et al. (2018; $\ln(\Lambda) = 3.9$). There is little existing work that numerically tests the analytic prediction for the magnitude of the drag force for a uniform medium at high Mach numbers, so it is difficult to tell if the fact that the drag force at high $\mathcal{M}$ is up to 50% higher than in the predicted analytic case for constant Coulomb logarithm is a generic result or not. Beckmann et al. (2017) do have a simulation at $\mathcal{M} = 10$ and results possibly also show a higher drag force at such high Mach numbers than the analytic prediction, but do not explore this further. This might be an interesting phenomenon to be explored in future work.

Firmer conclusions are difficult to draw due to the large scatter, induced by the BH crossing its own wake during repeated passages of the box.